\documentclass[amsmath,amssymb,aps,prb,twocolumn,floatfix,superscriptaddress]{revtex4-1} 
\usepackage{graphicx}
\usepackage{stmaryrd}
\usepackage{dcolumn}
\usepackage{bm}
\usepackage{xcolor}
\usepackage{multirow}
\usepackage{times}
\usepackage{chngcntr}
\usepackage{hyperref}
\usepackage{lineno} 

\definecolor{codegreen}{rgb}{0,0.6,0}
\definecolor{codegray}{rgb}{0.5,0.5,0.5}
\definecolor{codepurple}{rgb}{0.58,0,0.82}
\definecolor{backcolour}{rgb}{0.95,0.95,0.92}

\usepackage{listings}

\lstdefinestyle{mystyle}{
    backgroundcolor=\color{backcolour},   
    commentstyle=\color{codegreen},
    keywordstyle=\color{magenta},
    numberstyle=\tiny\color{codegray},
    stringstyle=\color{codepurple},
    basicstyle=\ttfamily\footnotesize,
    breakatwhitespace=false,         
    breaklines=true,                 
    captionpos=b,                    
    keepspaces=true,                 
    numbers=left,                    
    numbersep=5pt,                  
    showspaces=false,                
    showstringspaces=false,
    showtabs=false,                  
    tabsize=2
}

\newcommand{\vk}{\ensuremath{\vc{k}}}

\renewcommand{\vr}{\ensuremath{\vc{r}}}

\newcommand{\Ghost}{\ensuremath{G^{\rm host}}}
\newcommand{\Gimp}{\ensuremath{G^{\rm imp}}}

\newcommand{\DV}{\ensuremath{\,\Delta V}}

\renewcommand{\Im}{\mathrm{Im}}

\newcommand{\Tr}{\mathrm{Tr}}

\newcommand{\ST}{Sb$_2$Te$_3$}

\newcommand{\imp}{\ensuremath{\rm imp}}

\newcommand{\vc}[1]{\ensuremath{\boldsymbol{#1}}}

\newcommand{\beq}{\begin{equation}}
\newcommand{\eeq}{\end{equation}}
\newcommand{\eq}[1]{Eq.~(\ref{#1})}
\newcommand{\lbl}[1]{\label{#1}}

\newcommand{\code}[1]{{\normalfont\ttfamily #1}}

\newcommand{\tit}[1]{\textit{#1}.}
\newcommand{\vol}[1]{\textbf{#1}}
\renewcommand{\doi}[1]{{ doi:#1}}

\newcommand{\cita}[1]{\cite{#1}}
\usepackage[normalem]{ulem} 

\newcommand{\addition}[1]{{#1}} 
\newcommand{\replace}[2]{{#1}} 

\begin{document}

\title{The AiiDA-KKR plugin and its application to high-throughput impurity embedding into a topological insulator}


\author{Philipp R\"u{\ss}mann} 
\email[Corresponding author: ]{p.ruessmann@fz-juelich.de}
\affiliation{Peter Gr\"unberg Institut and Institute for Advanced Simulation, 
	Forschungszentrum J\"ulich and JARA, D-52425 J\"ulich, Germany}
	
\author{Fabian Bertoldo}
\affiliation{Peter Gr\"unberg Institut and Institute for Advanced Simulation, 
	Forschungszentrum J\"ulich and JARA, D-52425 J\"ulich, Germany}
	
\author{Stefan Bl\"ugel} 
\affiliation{Peter Gr\"unberg Institut and Institute for Advanced Simulation, 
	Forschungszentrum J\"ulich and JARA, D-52425 J\"ulich, Germany}


\begin{abstract}
	The ever increasing availability of supercomputing resources led computer-based materials science into a new era of high-throughput calculations.
	Recently, Pizzi \textit{et al.}\ [Comp.\ Mat.\ Sci.\ \textbf{111}, 218 (2016)] introduced the AiiDA framework that provides a way to automate calculations while allowing to store the full provenance of complex workflows in a database.
	We present the development of the AiiDA-KKR plugin that allows to perform a large number of \textit{ab initio} impurity embedding calculations based on the relativistic full-potential Korringa-Kohn-Rostoker Green function method.
	The capabilities of the AiiDA-KKR plugin are demonstrated with the calculation of several thousand impurities embedded into the prototypical topological insulator \ST.
	The results are collected in the JuDiT database which we use to investigate chemical trends as well as Fermi level and layer dependence of physical properties of impurities. 
	This includes the study of spin moments, the impurity's tendency to form in-gap states or its effect on the charge doping of the host-crystal.
	These properties depend on the detailed electronic structure of the impurity embedded into the host crystal which highlights the need for \textit{ab initio} calculations in order to get accurate predictions.
\end{abstract}

\maketitle


\section{Motivation}

In recent years computer-driven materials design has become  increasingly important in the field of materials science. 
The ever increasing availability of supercomputing resources opened up new possibilities towards data-driven condensed matter research.
Apart from large collections of crystal structure information \cite{PaulingFile,ICSD,COD,Nomad}, fully integrated frameworks of tools and databases have arisen that allow for high-throughput investigations using a huge amount of, mainly, density-functional-theory-based calculations \cite{ASE,AFLOW,MP,Pizzi2016}. Here, we present the {AiiDA-KKR} plugin \cite{aiida-kkr} which connects our full-potential relativistic Korringa-Kohn-Rostoker Green function (KKR) method \cita{jukkr} to the AiiDA (Automated Interactive Infrastructure and Database for Computational Science) framework \cita{Pizzi2016,aiidaurl}. 

The AiiDA infrastructure implements the FAIR principle \cita{FAIR} of findable, accessible, interoperable and reusable data sharing which provides a flexible plugin-based python environment. Through a common interface, different density-functional theory codes \cita{aiida-plugin-repository} can even be used in combination to exploit the strengths of different implementations and realise multi-code workflows within the same framework.
The KKR method is an all-electron implementation of density functional theory \addition{that allows accurate electronic structure calculations \cita{Ebert2011}, the extraction of magnetic response functions \cita{Liechtenstein1987} or gives access to transport properties \cita{Heers2011,Long2014,Zimmermann2016}}.
\replace{Its}{One of the advantages of the} Green function formulation \replace{allows, for instance,}{of the KKR method, which we focus on in this work, is} the efficient treatment of defective systems (i.e.\ systems that contain defects and impurities) which can be very expensive to treat with wavefunction-based methods that often require very large supercells for this task.
Our newly developed AiiDA-KKR plugin is used to perform a large number of impurity embedding calculations into the prototypical topological insulator \ST. 

Topological insulators (TIs) have been the center of attention in solid state research since their extraordinary physical properties, that lead to topologically protected surface states, have been discovered \cita{Hasan2010}. In the past decade the field around topological materials has evolved steadily and now aims at functionalizing TI materials by interfacing them with other states of matter. For instance, realizing the quantum anomalous Hall (QAH) \cita{Yu2010,Chang2013} insulator state or Majorana zero modes \cita{DasSarma2015}, that might lead to topological qubits, is pursued. The former requires to combine the topological band structures of TIs with magnetism while the latter needs interfacing topological materials with superconductors. 
Controlling the interface and understanding the effect defects and imperfections have remains a major challenge in this field to this day.

Apart from the development of the AiiDA-KKR plugin, the outcome of this study is the {JuDiT} database (\textbf{Jü}lich \textbf{D}atabase of \textbf{i}mpurities embedded into a \textbf{T}opological insulator) of physical properties of impurities embedded into the surface of \ST. We study their tendency for charge doping (i.e.\ to introduce $p$- or $n$-doping), their impurity magnetic moments and their density of states (DOS). This collection of impurity properties allows to uncover chemical trends and can help to optimize the next generation of TI-based materials in the future. 
In particular, we investigate the layer and Fermi level dependence of the spin moment and find that Mo$_\mathrm{Sb}$ defects show a high spin moment while introducing only a small charge doping, which is more than 5 times smaller than for magnetic $3d$ impurities. Furthermore, we find the Mo defect to be a good candidate for future applications since no impurity resonance appears in the bulk band gap region. An impurity resonance would otherwise lead to higher scattering rates of topological surface state electrons off this defect \cita{Ruessmann2017} and a higher probability to produce impurity bands in the gap.

The paper is structured as follows. First, the theoretical setting of the impurity embedding problem within the KKR method is introduced in section~\ref{sec:compdetails}. Then the AiiDA-KKR package is presented (section~\ref{sec:aiida-kkr}) where the calculation and workflow plugins, that are implemented in AiiDA-KKR, are discussed. Afterwards the developments are showcased at the example of high-throughput impurity embedding into the topological insulator material \ST\ in section~\ref{sec:impsST}. Finally, section~\ref{sec:conclusion} concludes with a summary. 


\section{\textit{Ab initio} impurity embedding \lbl{sec:compdetails}}

One of the advantages that arises from the Green function formulation of the KKR method lies in its ability to include impurities efficiently into crystalline solids \cita{Ebert2011,Bauer2013}. This is achieved making use of the {Dyson equation}
\beq
    \Gimp = \Ghost + \Ghost \DV \Gimp
\lbl{eq:Dyson}
\eeq
where $\Ghost$ is the Green function of the crystalline host system, 
$\DV = V^{\imp}-V^{\mathrm{host}}$ is the difference in the potential introduced due to the presence of the impurity and $\Gimp$ is the Green function that describes the impurity embedded into the periodic host crystal. It is important to mention that the change in the potential $\DV$ occurs only in a small region around the impurity which is why the Dyson equation can be solved in a small real space region around the impurity site. This \emph{impurity cluster} contains a few neighboring shells of host atoms that are necessary to properly treat the charge screening of the impurity by the neighboring host atoms. It is worthwhile noting that $\Gimp$ contains the complete information on physical properties like the density of states which is computed as $\rho(\vr; E)=-\frac{1}{\pi}\Im \Tr \Gimp(\vr,\vr; E)$ (the trace is implied over spin-, atom- and orbital momentum degrees of freedom of the Green function).
\addition{This impurity embedding scheme assumes a single impurity embedded into the infinite host crystal and therefore locally breaks the translational invariance. Such a calculation is typically done for the dilute limit of defects where the Fermi level is assumed to be fixed by the host crystal. The embedded defect is then allowed to transfer charge to and from the surrounding atoms of the host crystal that are included in the impurity cluster. 
The collective effect of changing Fermi level can then be included by using the host Green function computed for shifted Fermi level which can affect the impurity's charge state and may strongly affect the crystal's overall charge doping \cita{VandeWalle2004}.}

\replace{An}{In summary, an} impurity embedding calculation in the KKR formalism therefore consists of (i) calculating $\Ghost$ (from a converged host calculation), (ii) creating $\DV$, (iii) performing a self-consistent field cycle (scf) to converge $\Gimp$.


\section{The A\lowercase{ii}DA-KKR plugin \lbl{sec:aiida-kkr}}

The AiiDA framework \cita{Pizzi2016, aiidaurl} is a python package that allows to provide a code agnostic interface for different \textit{ab initio} codes which enables the automation of calculations. Simultaneously, inputs and outputs of complex sequences of calculations are automatically saved in a database \cita{postgresql}. This ensures the reproducibility of all results due to the stored provenance, which consist of nodes with directed connections, in the database. 
AiiDA implements the ADES model \cita{Pizzi2016} which provides a common layer of data structures that are used by the plugins of different codes \cita{aiida-plugin-repository}. This enables workflows that use multiple codes, which allows interoperability and exploitation of the individual strengths of different implementations. 

To enable high-throughput KKR calculations with AiiDA we developed the open source AiiDA-KKR python plugin \cita{aiida-kkr} that provides a set of calculations and workflows (i.e.\ complex sequences of calculations) and some accompanying tools. A detailed and up-to-date description of the input and output structure of the individual calculations, workflows and tools of AiiDA-KKR is included in its online documentation \cita{aiida-kkr-readthedocs} where additionally examples for the usage of the plugin are given. In the following only a brief overview of the features implemented in the AiiDA-KKR plugin is given.

\subsection{Calculation plugins}

\begin{figure*}[htb]
    \centering
    \includegraphics[width=17cm]{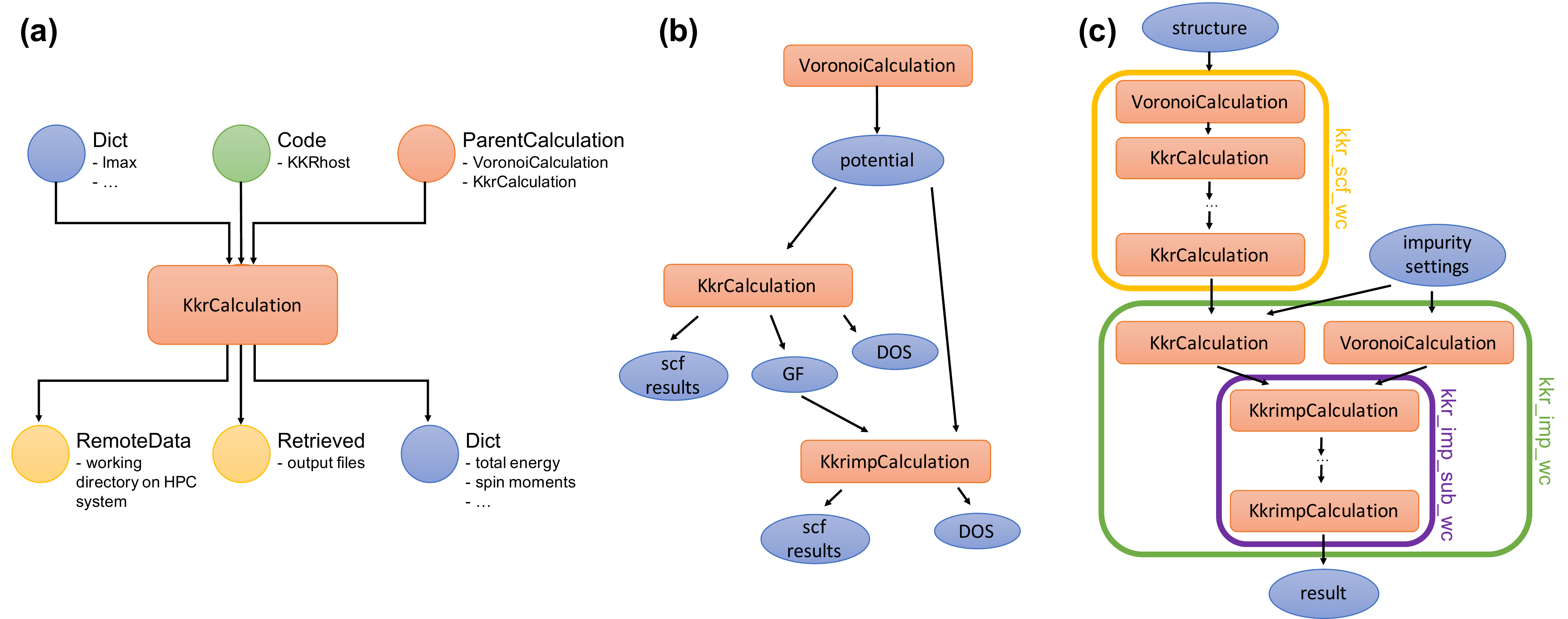}
    \caption{(a) Database structure of a typical KKR calculation performed with the AiiDA-KKR plugin. The node types and typical contents of the nodes are indicated and the arrows between nodes indicate the acyclicity of the graph. \replace{(b) Graph of a complete impurity-embedding calculation containing the full provenance including $\Ghost$-writeout, impurity potential setup (i.e.\ construction of $\DV$), and self-consistency convergence in multiple steps. While the complete sequence of calculations can be quite complex the full provenance from the final result over all intermediate steps up to the initial structure can be traced back in the provenance graph.}{(b) Dependencies between the outputs and inputs of different calculations supported by the AiiDA-KKR plugin. (c) Simplified view of a chain of calculations with an indication which parts are automated by different nested workflows of AiiDA-KKR.}}
    \label{fig:kkrcalc}
\end{figure*}

Each calculation plugin comes with the functionality to create code-specific input files from AiiDA objects (e.g. \code{StructureData} objects that contain the structural information of a system) and a parser that is able to parse the retrieved output files. The conversion of AiiDA \code{Dict} objects to the input file needed by the JuKKR code is facilitated with a python class called \code{kkrparams} \cita{masci-tools} that also contains methods to verify the consistency of input parameter and write the input file. This class also knows about KKR-specific features like dealing with alloys within the coherent potential approximation (CPA) \cita{Ebert2011}. Running a KKR calculation through AiiDA creates an acyclic directed graph in the database which is shown schematically for a \code{KkrCalculation} in Fig.~\ref{fig:kkrcalc}(a). It can be seen that a calculation requires a few input nodes (input parameter \code{Dict} node, a \code{Code} node, and a \code{ParentCalculation} node), which are all AiiDA objects stored in the AiiDA database. The output files are retrieved once the calculation finished and parsed to extract output parameters, that are then stored as a \code{Dict} node in the database. It it worthwhile noting that the AiiDA daemon takes care of automatically storing the resulting nodes with their directed connections to the AiiDA database. This eventually results in a complex graph as it is shown \replace{exemplary}{schematically} in \replace{Fig.~\ref{fig:kkrcalc}(b)}{Figs.~\ref{fig:kkrcalc}(b,c)} for a complete impurity-embedding sequence including all steps outlined in section \ref{sec:compdetails}. 
The full provenance of a complex procedure of calculations is stored in the database which allows to reproduce all inputs and the intermediate steps that have been performed to arrive at the final result.
In the following the different calculations provided in the AiiDA-KKR plugin \cite{aiida-kkr} are briefly discussed.

\begin{itemize}
    \item {The \code{VoronoiCalculation} plugin} allows to use the \textit{voronoi} code of the JuKKR package \cita{jukkr} which constructs the shape functions \cita{Stefanou1990, Stefanou1991} needed for the full-potential treatment and generates starting potentials. The graph of a \code{VoronoiCalculation} looks similar to the one shown in Fig.~\ref{fig:kkrcalc}(a) except that the \code{ParentCalculation} input node is replaced by an AiiDA \code{StructureData} node that contains all structural information on the crystal (e.g. lattice constant, atom positions and kinds) \cita{aiidaurl}. 
    \item The \code{KkrCalculation} plugin provides an interface to the \textit{KKRhost} code of JuKKR \cita{jukkr} which allows to perform self-consistency (scf), density of states (DOS), bandstructure and additional postprocessing calculations (e.g.\ calculation of Heisenberg exchange interaction parameters \cita{Liechtenstein1987}).
    \item The \code{KkrimpCalculation} plugin connects AiiDA to the \textit{KKRimp} code of the JuKKR package that solves the Dyson equation for impurity embedding (\eq{eq:Dyson}). This calculation needs in addition to the usual inputs (input parameter (\code{Dict}) or \code{ParentCalculation} node) the host Green function in the impurity cluster region ($\Ghost$) which is written out with a special post-processing run-mode of the KKRhost program using a \code{KkrCalculation}. The \code{KkrimpCalculation} can then be used to perform electronic structure calculations for the impurity problem. 
\end{itemize}

\subsection{Workflow plugins}

In addition to the calculation plugins, the AiiDA-KKR package provides some workflows that automate complex sequences of \textit{voronoi}, \textit{KKRhost} and \textit{KKRimp} calculations. The workflows contained in AiiDA-KKR have been developed in a modular way and build upon each other. \addition{This is illustrated in Fig.~\ref{fig:kkrcalc}(c) where the relation of the \code{kkr\_imp\_wc} and \code{kkr\_imp\_sub\_wc} workflows is shown.} Internally these are AiiDA \code{WorkChain}s which is indicated by the \code{\_wc} ending in the names of the workflows. For an in-depth discussion of the input and output structure and their usage we refer to the online documentation \cita{aiida-kkr-readthedocs}. Here we restrict our discussion to a short overview of the workflows of AiiDA-KKR.

\begin{itemize}
    \item The \code{kkr\_dos\_wc} workflow conveniently wraps around a \code{KkrCalculation} and provides the necessary inputs to perform a DOS calculation. Additionally the output is parsed and the output DOS data is stored as an array in the database which allows easy access and plotting of the output DOS.
    \item The \code{voro\_start\_wc} workflow wraps the \code{VoronoiCalculation} and performs some additional verification of input structure and KKR-specific parameters in order to make sure the chosen starting setting is reasonable. One of the checks performed automatically within \code{voro\_start\_wc} makes use of the \code{kkr\_dos\_wc} workflow.
    \item The \code{kkr\_scf\_wc} workflow builds upon the \code{voro\_start\_wc} workflow and a sophisticated series of \code{KkrCalculations} which is intended to reach convergence of a given host system reliably. This makes sure the starting setup is reasonable before the potential is pre-converged until finally convergence with higher accuracy settings is pursued. 
    \item The \code{gf\_writeout\_wc} workflow takes care of setting the necessary options for a \code{KkrCalculation} in order to write out $\Ghost$ in preparation of the impurity embedding step. For a given impurity position and screening cluster size the host Green function can be reused for several impurities that respect this embedding geometry.
    \item The \code{kkr\_imp\_sub\_wc} workflow performs the self consistency cycle of \textit{KKRimp} calculations similar to the logic implemented in the \code{kkr\_scf\_wc} workflow. It includes features that deal with possible convergence problems automatically.
    \item The \code{kkr\_imp\_wc} workflow combines the \code{voro\_start\_wc}, \code{gf\_writeout\_wc} and \code{kkr\_imp\_sub\_wc} steps for the impurity problem. This allows to conveniently start with a converged host calculation and the information on the impurity (e.g.\ its position in the host crystal, the size of the screening cluster) which defines the problem given by \eq{eq:Dyson} completely. 
    \item The \code{kkr\_imp\_dos\_wc} workflow gives an easier access to calculate the DOS of an impurity embedded into a host crystal. 
\end{itemize}

\subsection{Tools}

\begin{figure}[htb]
    \centering
    \includegraphics[width=8.5cm]{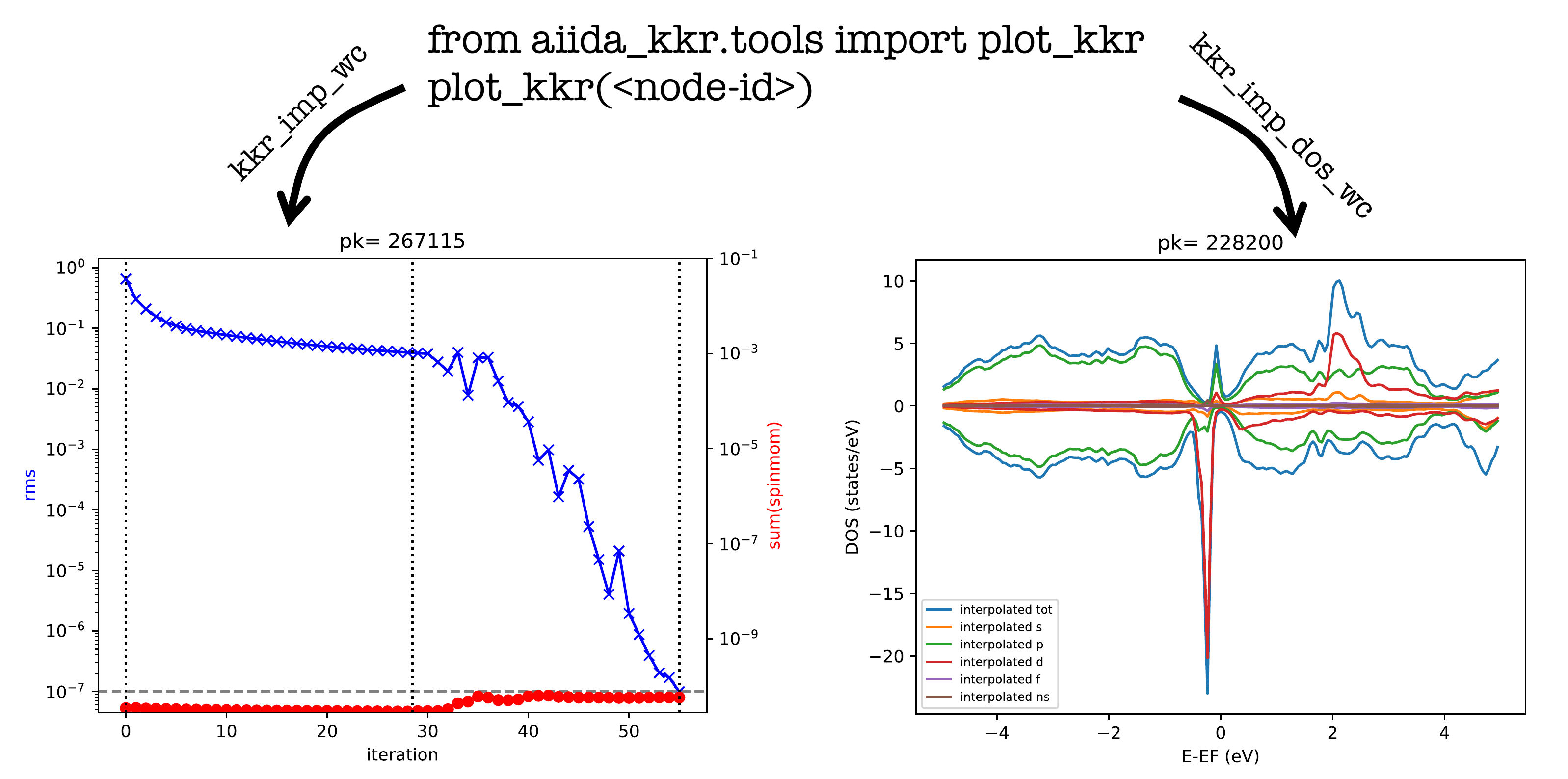}
    \caption{Illustration of typical plots generated with the \code{plot\_kkr} tool of AiiDA-KKR (left: convergence behavior of a \code{kkr\_imp\_wc} workflow, right: DOS output of a \code{kkr\_imp\_dos\_wc} workflow).}
    \label{fig:plot_kkr}
\end{figure}

Apart from tools that are used internally within the calculation and workflow plugins (e.g.\ to prepare the real space screening cluster), the AiiDA-KKR plugin contains a plotting tool called \code{plot\_kkr}. This tool takes a node identifier (the AiiDA node instance, its \textit{pk} or \textit{uuid}) or even a list of nodes and plots a standard, yet fully customisable, plot of the respective data. For example, a typical self-consistency workflow for an impurity calculation (\code{kkr\_imp\_wc} node) as input to \code{plot\_kkr} will produce a plot of the convergence behavior with the scf iteration number or an instance of a \code{kkr\_imp\_dos\_wc} workflow will produce the orbital-resolved plot of the total DOS in the impurity cluster. Examples of such plots are given in Fig.~\ref{fig:plot_kkr}. The tool conveniently abstracts away the need to extract the relevant data from the nodes in the database and provides a straightforward way to create commonly used plots.

\subsection{\addition{Example usage: Ag defect embedded into bulk Cu}}

\addition{
To illustrate how AiiDA-KKR's python interface facilitates complex density functional calculations we demonstrate the use of the top-level workflows \code{kkr\_scf\_wc} and \code{kkr\_imp\_wc} at the example of an Ag impurity embedded into bulk fcc Cu. Solving this problem requires setting up the starting potential, converging the bulk electronic structure, writing out the host's Green function for the impurity embedding and finally performing the impurity embedding step as sketched in Figs.~\ref{fig:kkrcalc}(b,c). The AiiDA-KKR plugin conveniently automates this complex series of tasks. The following code snippet illustrates how the self-consistent calculation for the Cu bulk can be submitted to the AiiDA daemon that takes care of orchestrating the necessary sequence of \code{VoronoiCalculation} and \code{KkrCalculation}s.
}

\begin{figure}[htb]
    \centering
    \includegraphics[width=7.5cm]{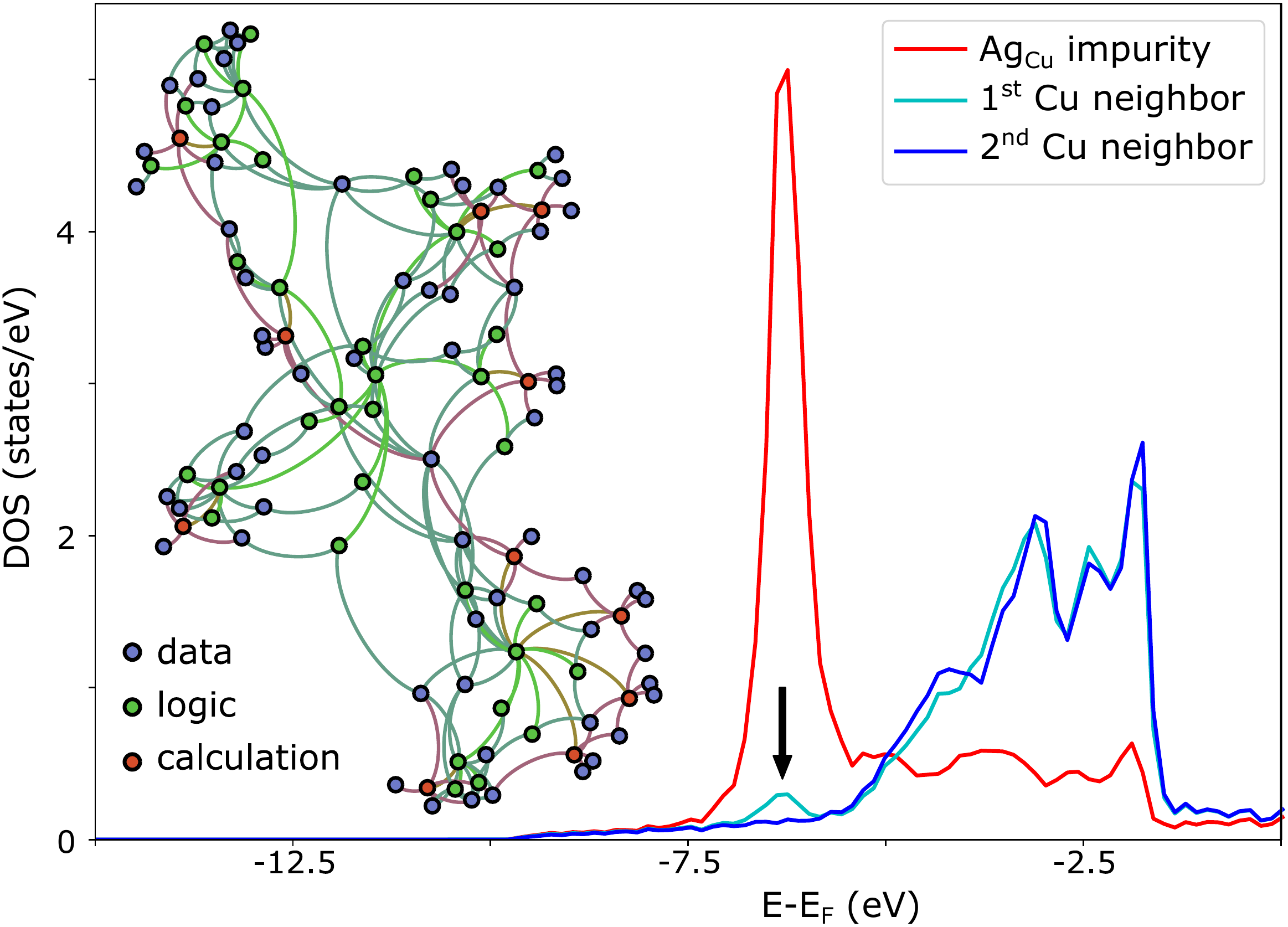}
    \caption{\addition{Density of states of an $\mathrm{Ag}_{\mathrm{Cu}}$ impurity and its surrounding Cu neighbors. The black arrow highlights the hybridization-induced states in the first Cu neighbor arising from the interaction with the $d$-resonance of the Ag defect. The inset (left) visualizes the database structure of the complete $\mathrm{Ag}_{\mathrm{Cu}}$ calculation which shows the complex relation between data nodes (structure input node, results \code{Dict} nodes etc.), logical nodes (e.g.\ different steps in \code{kkr\_scf\_wc} or \code{kkr\_imp\_wc} workflows) and calculations (\code{VoronoiCalculation}, \code{KkrCalculation} and \code{KkrimpCalculation}).}}
    \label{fig:AgCu}
\end{figure}

\begin{lstlisting}[basicstyle=\linespread{1}\scriptsize, language=Python, caption=\addition{Submission of a self-consistency workflow for bulk Cu. The crystal structure is specified with the \code{structure} input and basic KKR-specific parameters, that control the accuracy of the calculation, are set in the \code{calc\_parameters} input node. Here \code{submit}, \code{Dict} and \code{Code} are basic AiiDA methods and classes.}, label=lst:bulkscf]
scf = submit(kkr_scf_wc, 
    voronoi=Code.get_from_string('voronoi@localhost'),
    kkr=Code.get_from_string('kkrhost@localhost'),
    calc_parameters=Dict(dict={'LMAX': 2, 'NSPIN': 1,
                               'RMAX': 7., 'GMAX': 65.)),
    structure=Cu_bulk_structure)
\end{lstlisting}
\addition{
Once the calculation for bulk fcc Cu finishes, the impurity embedding step can be done. This is equally simple with AiiDA-KKR's python interface as the following code snippet demonstrates.
}
\begin{lstlisting}[basicstyle=\linespread{1}\scriptsize, language=Python, caption=\addition{Submission of the complete impurity embedding workflow that starts from the converged calculation of the Cu bulk system (called \code{scf} in Listing~\ref{lst:bulkscf}). The impurity is specified via its nuclear charge, the position in the host crystal and the radius of the impurity cluster via the \code{impurity\_info} input node.}]
submit(kkr_imp_wc,
    impurity_info=Dict(dict={'ilayer_center': 0,
                             'Zimp': 47, 'Rcut': 4.0}),
    kkr=Code.get_from_string('kkrhost@localhost'),
    voronoi=Code.get_from_string('voronoi@localhost'), 
    kkrimp=Code.get_from_string('kkrimp@localhost'),
    remote_data_host=scf.outputs.last_RemoteData)
\end{lstlisting}
\addition{
In this example we define the Ag impurity which replaces a Cu atom and include a screening cluster around the impurity of $4\,\mathrm{\AA}$ which contains the first two shells of Cu neighbors. From the converged calculation we can investigate the impurity DOS which is shown in Fig.~\ref{fig:AgCu}. It can be seen that embedding an Ag atom in the Cu crystal locally changes the electronic structure around the defect. This becomes evident in the appearance of a peak in the nearest neighbor Cu DOS at the position in energy where the Ag atoms has a resonance its DOS (highlighted by the black arrow). This resonance disappears already in the second Cu neighbor which shows the localized character of the impurity state. In addition, the inset in Fig.~\ref{fig:AgCu} visualizes the resulting database structure of such a series of calculations (bulk scf, impurity scf, impurity DOS) where each node in the graph is an entry in the AiiDA database. The number of nodes and their connections highlight the complexity of the impurity embedding task that is conveniently taken over by the AiiDA-KKR plugin.
}

\section{\replace{\lowercase{imp}Dat:TI}{J\lowercase{u}D\lowercase{i}T} -- A database for impurities embedded into a TI \lbl{sec:impsST}}

We apply the AiiDA-KKR plugin to embed a large number of impurities into the topological insulator \ST. The resulting {\mbox{JuDiT}} database \cita{doi-database} comes with a webinterface \cita{website-impDat} for convenient access to the included data. 
The following analysis shows some of the physical insights obtained through this study. Our data analysis does not aim at being comprehensive but is intended to showcase the usefulness of our application. A future data-driven study might give additional insights but is beyond the scope of this work.

\subsection{\ST\ host system and computational details}

\begin{figure}[htb]
    \centering
    \includegraphics[width=8.5cm]{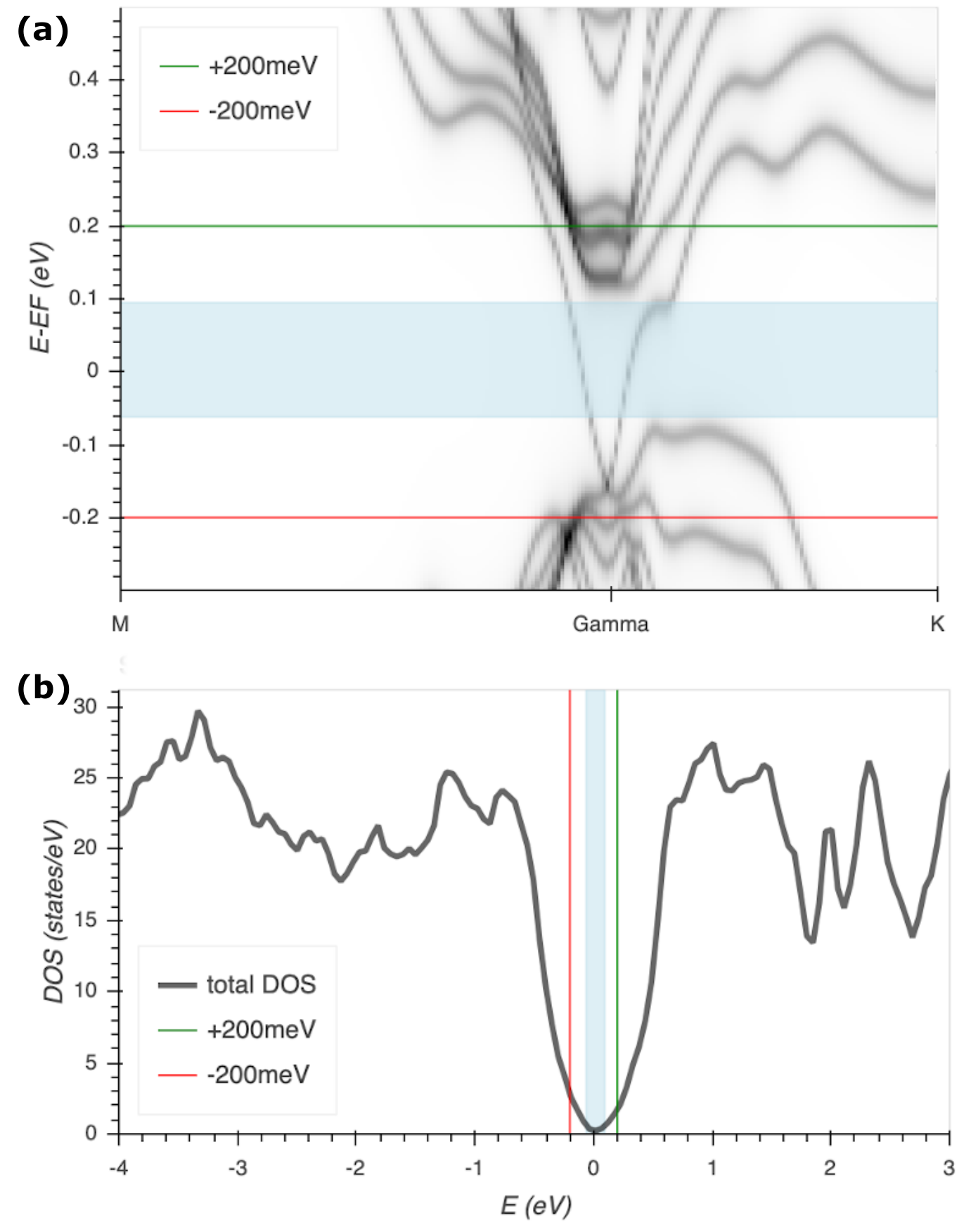}
    \caption{Band structure in terms of the Bloch spectral function (a) and density of states (b) of the 6 quintuple-layer thick Sb$_2$Te$_3$ host crystal. Indicated are the bulk band-gap region (light blue region) as well as the in the considered Fermi level shifts (red and green lines) that are used to simulate the effect of $p$- and $n$-doping in the topological insulator host material.}
    \label{fig:bandstruc_dos_host}
\end{figure}

For the JuDiT database we considered substitutional defects (denoted by $X_Y$ for impurity $X$ replacing host atom $Y$) in the first 3 quintuple layers (QL) of a 6 QL thick film of \ST. This allows to study the influence of the topological surface state, that is mainly located in the 1st QL, on impurity properties. The band structure and density of states of the clean host system are shown in Fig.~\ref{fig:bandstruc_dos_host}. In order to take into consideration doping of the host material, we investigated three possible positions of the Fermi level ($E_F$ located in the valence band (VB) and in the conduction band (CB), as well as $E_F$ in the bulk band gap). The shifted positions of the Fermi level are highlighted in Fig.~\ref{fig:bandstruc_dos_host} with red ($E_F$ in VB) and green ($E_F$ in CB) lines.
For the impurity embedding we neglected structural relaxations of the atoms around the impurities but included the first 3-4 shells of neighbors (containing 21-27 atoms and empty cells in the impurity cluster) within a radius of $4.8\,\mathrm{\AA}$ around the impurities. We used the exact description of the atomic cells \cita{Stefanou1990,Stefanou1991} and the local spin density approximation \cita{Vosko1980} (LSDA) for the exchange correlation functional. A cutoff for the angular expansion of $\ell_{\mathrm{max}}=3$ was chosen and corrections for the truncation error using Lloyd's formula have been applied \cita{Zeller2004}. Relativistic corrections arising within the scalar-relativistic approximations as well as spin-orbit coupling have been taken into account fully self-consistently for both host and impurity calculations. \addition{This approach has been applied successfully in the past to study magnetic and non-magnetic defects in topological insulators where a good agreement between our theoretical predictions and different experimental observations was verified \cita{paperfocusing,paper-Barla,Ruessmann2017,Peixoto2019}.}

\subsection{Physical properties of impurities embedded into \ST}

We start our analysis with an overview of the contents of the JuDiT database. In total more than 2100 impurities have been embedded self-consistently into the \ST\ host system. For each impurity we computed physical properties like the spin and orbital moments, the impurity's DOS as well as the tendency to show impurity resonance in the region of the bulk band gap.
Furthermore, we analyzed the charge doping introduced by the defect which we define as
\begin{equation}
    \Delta n^{\mathrm{imp}}= \bigl(n^{\mathrm{imp}}-Z^{\mathrm{imp}}\bigr) - \bigl(n^{\mathrm{host}}-Z^{\mathrm{host}}\bigr),
    \label{eq:Deltanimp}
\end{equation}
where $n^{\mathrm{imp}}$ ($n^{\mathrm{host}}$) are the electron densities for the impurity (host) atom embedded into the surrounding host crystal integrated in the Voronoi cell of the atom. Here, $Z^{\mathrm{imp}}$ ($Z^{\mathrm{host}}$) is the nuclear charge of the impurity (host) atom. The expression in brackets in the rhs of \eq{eq:Deltanimp} therefore contain the information how much charge is transferred to/from the impurity. 

\begin{figure}[htb]
    \centering
    \includegraphics[width=8.5cm]{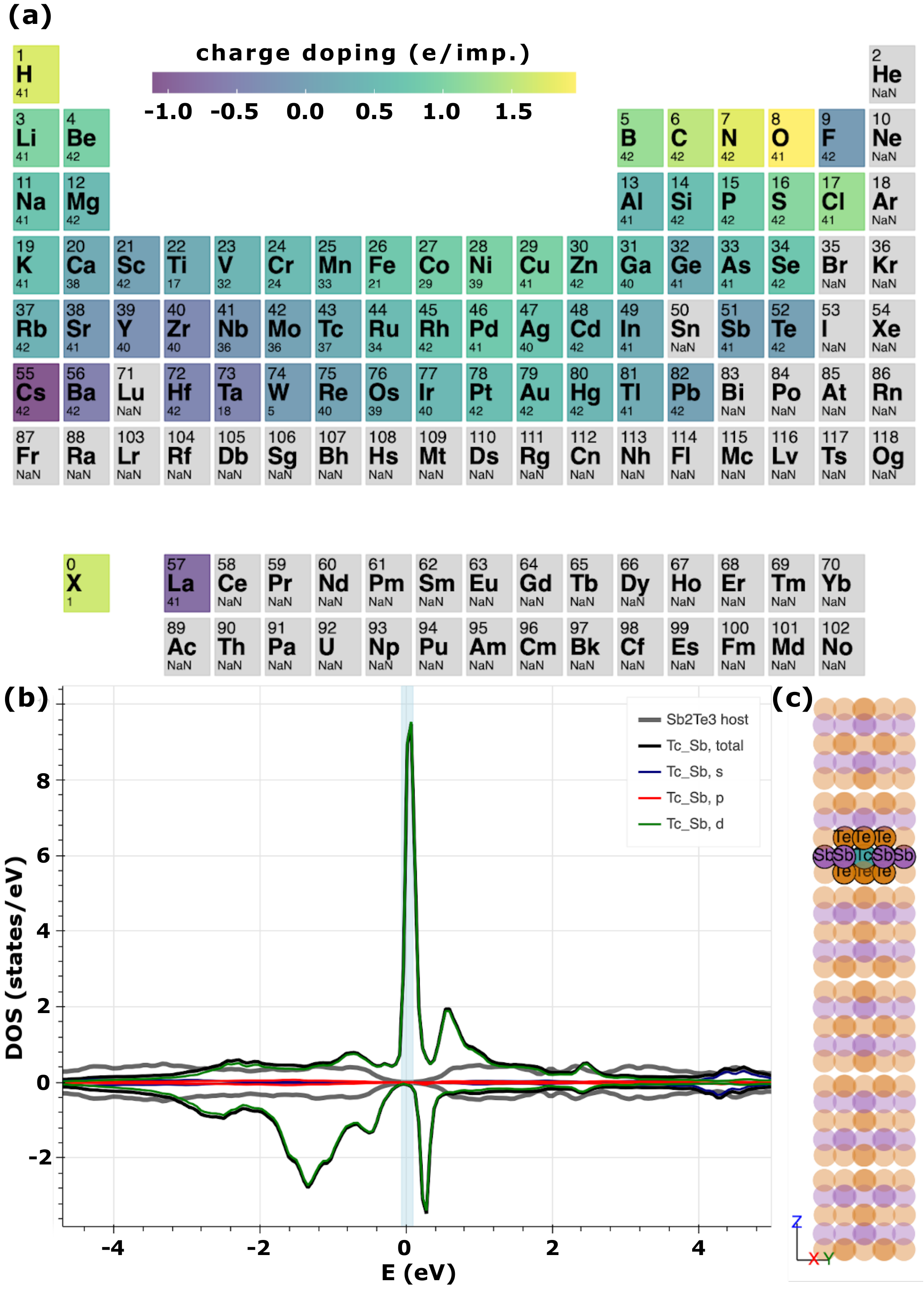}
    \caption{(a) Overview of the impurity-induced charge doping (averaged over all considered impurity configurations, given in electrons per impurity atom). (b) Impurity density of states for a Tc$_{\mathrm{Sb}}$ defect showing a resonance arising in the bulk band gap region (blue shaded region). Positive and negative values correspond to minority and majority states. (c) Location of the Tc$_{\mathrm{Sb}}$ impurity in the \ST\ host crystal. The full-colored spheres represent the atoms in the impurity cluster and the opaque atoms show the rest of the host crystal where $\DV=0$.}
    \label{fig:imp_results_overview}
\end{figure}

The results can conveniently be visualized with the \mbox{JuDiT} webinterface \cita{website-impDat}. The plots from Fig.~\ref{fig:bandstruc_dos_host} as well as the ones of Fig.~\ref{fig:imp_results_overview} have been created using the tools available there. Figure~\ref{fig:imp_results_overview}(a) displays the, over the different impurity sites averaged, charge doping given in electrons per impurity. The observed chemical trends partly fit the behavior of the Pauling electronegativity \cita{Pauling-electronegativity}. \addition{This causal relation is highlighted by a Pearson correlation coefficient of $0.5$ which is found between the induced charge density and the impurity's electronegativity (see Fig.~\ref{fig:app2.1} in the appendix for details).} The details of the bonding mechanism are however more subtle and can be quantified using \textit{ab initio} data. This reflects the intricate physics of the chemical bonding which was recently classified to be \emph{metavalent} for the \ST\ class of materials \cita{Yu2019}.

Each impurity in JuDiT also has a {detail page} where the complete output dictionary of the converged calculation is given and from where its provenance can be browsed. It also features a plot of the impurity DOS which is shown exemplary in Fig.~\ref{fig:imp_results_overview}(b) for a Tc$_\mathrm{Sb}$ impurity located in the fourth Sb layer from the surface. The position of the impurity in the \ST\ host crystal is visualized in Fig.~\ref{fig:imp_results_overview}(c). It can be seen that the $d$-states of the Tc atom are exchange- and crystal-field-split which results in a magnetic moment of the impurity and a resonance in the impurity DOS around the Fermi level consequently in the bulk band gap region (blue area in Fig.~\ref{fig:imp_results_overview}(b), see also Fig.~\ref{fig:bandstruc_dos_host}). This particular defect is therefore expected to lead to strongly scattering of the topological surface state electrons \cita{Ruessmann2017} which can induce significant back-scattering since the Tc defect is a resonant scatterer \emph{and} is magnetic with a spin moment of $1.86\,\mu_\mathrm{B}$.

In order to quantify the gap-filling nature of all considered defects, we define the number of states introduced by the defect in the bulk band gap region as
\begin{equation}
    n^{\mathrm{imp}}_{\mathrm{gap}}=\int\limits_{E_\mathrm{min}}^{E_\mathrm{max}}\mathrm{d}E \int\limits_{\Omega_\mathrm{Imp}}\mathrm{d}^3r \ \rho(\vr; E),
    \label{eq:nimp_dap}
\end{equation}
where $E_\mathrm{min}$ and $E_\mathrm{max}$ are the edges of the bulk band gap region (blue area in Fig.~\ref{fig:bandstruc_dos_host}), $\Omega_\mathrm{imp}$ is the Voronoi cell around the impurity atom and $\rho(\vr; E)$ denotes the charge density around the atom in the Voronoi cell. A high gap-filling value consequently signals that scattering off that particular impurity will be increased which could be detrimental to the desired transport properties of TI materials. This is especially the case if the impurity is magnetic and the $\vk\to-\vk$ backscattering channel reopens due to broken time reversal symmetry.

\subsection{Magnetic impurities}

\begin{figure*}[htb]
    \centering
    \includegraphics[width=17cm]{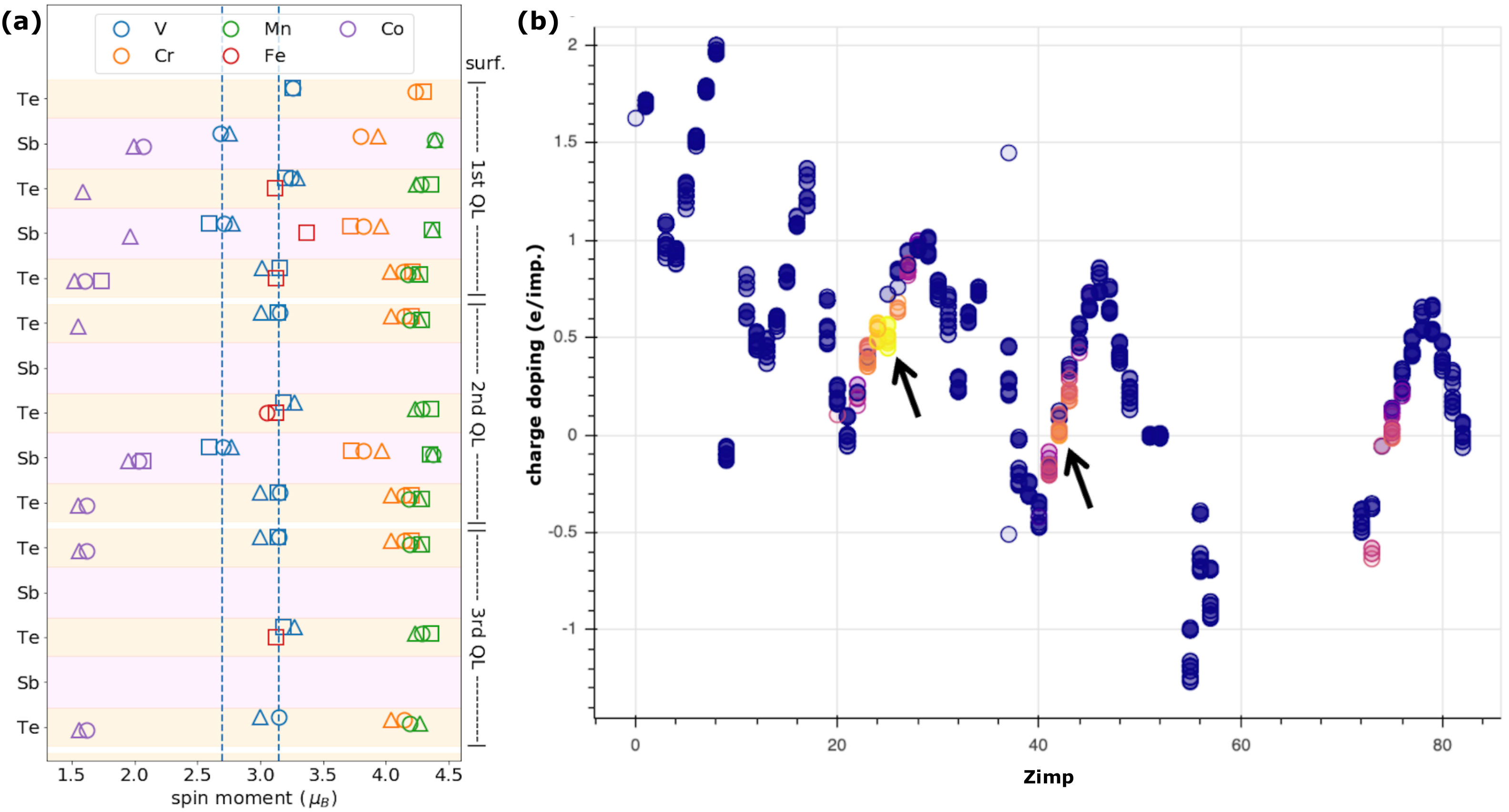}
    \caption{(a) Layer and Fermi level dependence ($E_F-200\,\mathrm{meV}$: $\triangle$, $E_F$: {\footnotesize $\bigcirc$}, $E_F+200\,\mathrm{meV}$: $\square$) of the spin moment of some $3d$-impurities. The blue dotted lines serve as guides to the eye to highlight the difference in the spin-moment for Te and Sb layers. (b) Charge doping (in units of added electrons per impurity) vs. impurity atomic charge of all considered impurities. The color of the data points in (b) refers to the magnitude of the spin moment (blue: non magnetic to yellow: $4.4\,\mu_B$). The arrows highlight magnetic $3d$ and $4d$ impurities which differ in their induced charge doping to the host crystal. The plot in (b) was generated using the {JuDiT} web interface \cita{website-impDat} that accompanies the publication and can be used to visualize and export the result for future studies.}
    \label{fig:layer_dependence}
\end{figure*}

We now focus our attention to magnetic defects which are found for some transition metal impurities. These systems are interesting in the context of realizing a robust quantum anomalous Hall phase. 
We start by investigating the layer and Fermi level dependence of the spin moment of $3d$ transition metal impurities shown in Fig.~\ref{fig:layer_dependence}(a). We see a gradual increase of the magnetic moment when going from V over Cr up to Mn dopants (blue orange and green symbols, respectively) before a subsequent decrease with Fe and Co impurities (red, violet) is observed. This behavior is expected from Hund's rule and reflects the, from V to Co, increasing filling of the $d$-shell (see also Fig.~\ref{fig:app1.1} in the appendix). 

The details of the impurity's electronic structure are largely determined by two factors: (i) the atomic nature of the impurity atom determining its number of electrons in the atomic configuration and (ii) the interaction with the surrounding atoms of the host crystal that affect the hybridization of the atomic states of the impurity with the host's band structure. This effect is seen in the layer dependence of the size of the spin-moment which is highlighted for V$_\mathrm{Sb}$ and V$_\mathrm{Te}$ defects with the blue dashed lines in Fig.~\ref{fig:layer_dependence}(a). We observe a higher spin-moment for V$_\mathrm{Te}$ compared to V$_\mathrm{Sb}$ which can be attributed to a larger charge transfer to the impurity in the Sb substitutional site compared to the Te substitutional site. The larger charge transfer to the impurity results in a higher filling of the V $d$-shell and therefore a higher spin moment which can also be seen for the Cr impurity. The same mechanism leads to a decrease in the spin moment that is found for Co$_\mathrm{Sb}$ defects compared to Co$_\mathrm{Te}$.

Investigating the dependence of the impurity spin moment on the Fermi level (different symbols in Fig.~\ref{fig:layer_dependence}(a)) reveals that the details of the hybridization with the electronic band structure of the host material can be controlled via the position of the Fermi level in the host material. 
A shift in the host's Fermi level can experimentally be achieved by appropriate doping with Bi$_\mathrm{Sb}$ impurities\cita{Kellner2015}. 
This change in the impurity moment has been seen previously both theoretical \cita{paper-Barla} as well as experimental \cita{Peixoto2019} and can be attributed to the competition between the impurity seeking charge neutrality and the strong change in the hybridization with the host's electronic structure with varying position of the Fermi level due to the presence of the bulk band gap in TI materials. This leads, for instance, to decreasing (increasing) spin moments for V$_\mathrm{Sb}$ (V$_\mathrm{Te}$) with increasing position of the Fermi level.

Overall we observe that for $V$ impurities the spread in the spin moment with the impurity's surrounding (i.e.\ its layer dependence) is twice as large as with varying Fermi level. On the contrary, for Mn defects the change in the spin moment in different layers and with varying Fermi level is always rather small. This results from the half-filling of the Mn $d$ orbital that make the spin moment relatively insensitive to small changes in the hybridization with the host's electronic structure. Nevertheless, the magnetic interactions among multiple magnetic impurity atoms can, even at small changes in the impurity hybridization, be strongly affected \cita{paper-Barla,Peixoto2019}.

Figure~\ref{fig:layer_dependence}(b) summarizes the charge doping $\Delta n^{\mathrm{imp}}$ for all defects included in JuDiT. The zigzag behavior with the impurity's core charge reflects the structure of the periodic table with its isoelectronic groups. \addition{This is verified by a Pearson correlation coefficient of $0.68$ between the impurity's group index and the induced charge doping (see Fig.~\ref{fig:app2.2} in the appendix for details).} Each data point \addition{in Fig.~\ref{fig:layer_dependence}(b)} is colored by the impurity's spin moment which shows that the maximum of the spin-moment is found for Mn and Fe impurities (yellow points). It can however be seen that the charge doping introduced by these $3d$ defects is fairly large which reflects the significant difference in electronegativity compared to the Sb and Te host atoms \cita{Pauling-electronegativity}.

Applying magnetic doping to achieve a robust quantum anomalous Hall phase needs to fulfill some boundary conditions in order to be feasible in experiments. In order to not tune the Fermi level out of the bulk band gap by magnetic doping, the induced charge doping should be as small as possible. At the same time magnetism is the key ingredient which calls for a sizable spin moment of the impurity. Furthermore, the magnetic impurity should not show a high DOS in the bulk band gap region to reduce the appearance of unwanted impurity bands with increasing magnetic doping. If we apply these conditions of a gap filling of $n^{\mathrm{imp}}_{\mathrm{gap}}<0.02\,\mathrm{e}$, a charge doping of $\Delta n^{\mathrm{imp}}<0.1\,\mathrm{e}$ and a spin-moment of $m_s>1.5\,\mathrm{\mu_B}$ we find that Mo$_\mathrm{Sb}$ defects meet all these criteria. Compared to the Te substitutional site, which also shows a small charge doping and high spin moment, the Sb substitution have a gap filling which is an order of magnitude smaller and could therefore be desirable. To be able to use Mo-dopants for the realization of the QAH state the magnetic exchange coupling between Mo atoms needs to be investigated further in the future. This is, however, beyond the scope of this work. The study of the exchange interactions will be especially interesting since Mo-doping of Bi$_2$Se$_3$ showed signatures of antiferromagnetic coupling \cita{MoAFMcalc} which could be possibly overcome by appropriate band structure and defect engineering as it was seen for Mn and Co doping of Bi$_2$Te$_3$ \cita{paper-Barla}. Here, additional co-doping with other defects could open another way to design TI-based materials for future applications \cita{Kim2017}.

\subsection{\addition{Effects of electron correlation calculated within LDA+U}}

\addition{
In the framework of density functional theory the effect of on-site Coulomb repulsion for localized $d$ or $f$ electrons can be included with the LDA+U method \cita{Ebert2003}. We applied this scheme for 153 transition metal defects in the JuDiT database.
For the parametrization of the correlations we used the U and J values calculated from the constrained random-phase approximation \cita{Sasioglu2011}. The values for the LDA+U parametrization used in this work for the transition metal impurities are given in the appendix. Generally, including correlations within the LDA+U method increases the exchange splitting in magnetic impurities. This is seen, for example, in the comparison of the DOS of $3d$ transition metal defects with and without U-corrections (see Fig.~\ref{fig:app1.2} in the appendix).
Adding correlation effects also changes other physical properties which manifests in an increase in the median value of the spin moment by 36\% (see Fig.~\ref{fig:app2.3} in the appendix) for the LDA+U calculations compared to the previously discussed LDA results. An even stronger effect is seen on the orbital moment where we find a decrease of 58\% in the median value for the subset of impurities where LDA+U calculations have been included.
These results indicate that local correlations can strongly affect the outcome and should be considered when comparing calculations and experimental results.
}

\subsection{\addition{Comparison to literature values}}

\addition{
In order to review the accuracy of our calculations and estimate possible shortcomings of our approach we attempt a comparison of physical properties contained in the JuDiT database to already published experimental and theoretical results. We focus on (i) the experimentally observed charge doping in Sb$_2$Te$_3$ and (ii) magnetic properties reported mainly for V and Cr doped (Bi,Sb)$_2$Te$_3$.

In experiments, Sb$_2$Te$_3$ is typically found to be $p$-doped which is associated to the abundance of intrinsic $\mathrm{Sb}_{\mathrm{Te}}$ anti site defects \cite{Lostak1989, Cava2013}. This trend is confirmed by results for the impurity induced charge doping in the JuDiT database which is found to be negative ($\approx-0.01\,e/\mathrm{imp.}$ for $p$-doped Sb$_2$Te$_3$).
Additionally, $\mathrm{Te}_{\mathrm{Sb}}$ defects show the same trend which is in line with the experimental observation that $n$-type doping is not realized in Sb$_2$Te$_3$ even under Te rich growing conditions \cita{Cava2013}.
However, the size of the charge doping that results from our calculations is rather small which might indicate that in strongly doped materials collective effects can lead to a more pronounced effect. This can, for instance, result from the long-ranged Coulomb interaction among charged defects that leads to a shift in the Fermi level of the whole crystal\cita{VandeWalle2004}. 

In the literature doping of Sb$_2$Te$_3$ is mostly studied in the context of possible realizations of the quantum anomalous Hall effect with magnetic dopants. In Table~\ref{tab:magmomLiterature} we collected some experimental (e.g.\ from XMCD data) and theoretical (e.g.\ DFT supercell calculations) results for the spin moment of transition metal doped Sb$_2$Te$_3$ compounds. We focus on the LDA results since most calculations in the literature do not report values including correlation effects.

We can see that there is a considerable spread in the reported values for the spin moment of different impurities. Taking this methodological variance into account we find a reasonable agreement with our data from the JuDiT database. We suspect that the slight overestimation of the spin moment for V and Cr is a result of the neglected structural relaxations around the defects. The appearance of a resonance in the impurity DOS around the Fermi level (see Figs.~\ref{fig:app1.1},\ref{fig:app1.2} in the appendix) could be removed by structural relaxations. In the spirit of the Jahn-Teller effect this could change the $d$-filling of the impurity and consequently lead to smaller spin moments for V and Cr defects. A future study focusing on this effect might give more insights into the effect of structural relaxations.
Some experiments \cita{Peixoto2016,Islam2018,Peixoto2019} additionally report on the measured impurity DOS for V and Cr defects from resonant photoemission spectroscopy or scanning tunneling spectroscopy. A comparison to  our calculated impurity DOS spectra shows good agreement for defects at the substitutional Sb site which is in line with the reasonable agreement of the spin moment reported in Table~\ref{tab:magmomLiterature}. 
}

\begin{table}[]
    \centering
    \begin{tabular}{c|c|c|c|c}
        host compound  & V & Cr & Mn & Fe \\\hline
        Sb$_2$Te$_3$, \textit{th.}\cita{Islam2018}                              & 2.0   & 2.9-3.0 & 4.0-4.1 & 3.2-3.3 \\
        (Bi$_{0.2}$Sb$_{0.8}$)$_2$Te$_3$ , \textit{th.}\cita{Peixoto2016}         & 2.6   & -       & -       & - \\
        Sb$_2$Te$_3$, \textit{th.}\cita{Peixoto2019}                             & 2.3   & 3.3     & -       & - \\
        Sb$_2$Te$_3$ , \textit{th.}\cita{Zhang2013}                                & -     & 3.1     & -       & 4.2 \\
        (Bi$_{0.1-0.3}$Sb$_{0.9-0.7}$)$_2$Te$_3$, \textit{ex.}\cita{Peixoto2019} & 2.6   & 3.8     & -       & - \\
        (Bi$_{0.1}$Sb$_{0.9}$)$_2$Te$_3$, \textit{ex.}\cita{Tcakaev2020}          & 2-2.4 & 2.3-3.2 & -       & - \\
        Sb$_2$Te$_3$, \textit{this work}                                       & 2.7   & 3.8     & 4.4     & 3.3
    \end{tabular}
    \caption{\addition{Spin moments in $\mu_B$ of $3d$ transition metal doped Sb$_2$Te$_3$ compounds. Both theoretical (\textit{th.}) and experimental (\textit{ex.}) data are collected in comparison to data from the JuDiT database (\textit{this work}). Theoretical results are only reported without correlation effects.}}
    \label{tab:magmomLiterature}
\end{table}


\section{Conclusions \label{sec:conclusion}}

In conclusion, we have developed the AiiDA-KKR plugin which is an open source python package that connects the JuKKR code family to the AiiDA framework. This allows to perform Korringa-Kohn-Rostoker Green function calculations in an automated high-throughput manner. We concentrated on the ability to perform \textit{ab initio} impurity embedding into the topological insulator \ST.

We considered several thousand different impurities embedded into the different layers of the \ST\ host crystal. This procedure allowed us to study the layer and Fermi level dependence of physical properties of defects. Specifically, we studied the chemical trends in terms of the impurity's charge doping, their tendency to create resonances in the bulk band gap and their magnetic properties. The results have been collected in the JuDiT database which is openly available and comes with online tools for data visualization and export. Throughout our analysis we have seen that the details of the electronic structure of an impurity embedded into a host crystal is very important. The hybridization of the impurity states with its surrounding plays a crucial role for its physical properties. This highlights the relevance and the need for our \textit{ab initio} calculations which provide predictive power.

In the future the AiiDA-KKR plugin in general and the resulting data of this study in particular can be used in broader high-throughput studies for quantum materials. The capabilities of AiiDA-KKR could be extended to further include the automated calculation of scattering \cita{Heers2011,Long2014,QPI-paper} and transport properties \cita{Zimmermann2016,Kosma2020} or to investigate magnetic exchange interactions \cita{Liechtenstein1987}.
Especially the vast space of combinations co-doping with other impurities introduces will be of interest in order to find ways to tune physical properties and engineer the behavior of TI-based materials.


\section*{Data Availability}

The data generated and analysed during the current study are available in the materialscloud archive\cite{Talirz2020,doi-database}. The JuDiT web interface\cite{website-impDat} is published at \href{https://www.materialscloud.org/discover}{https://www.materialscloud.org/discover}.
The figures of this work can be reproduced with the tools developed for the JuDiT web interface\cite{website-impDat}.

\section*{Code Availability}

\addition{
The sourcecode of the AiiDA-KKR plugin \cite{aiida-kkr} is published as open source software under the MIT license at \href{https://github.com/JuDFTteam/aiida-kkr}{https://github.com/JuDFTteam/aiida-kkr}.
The source code of the JuDiT web interface\cite{website-impDat} with its visualization capabilities is open source under the MIT license as well and can be found at \href{https://github.com/PhilippRue/JuDiT-discover-section}{https://github.com/PhilippRue/JuDiT-discover-section}.
}


\section*{Acknowledgements}

We acknowledge funding from the Priority Programme SPP-1666 Topological Insulators of the Deutsche Forschungsgemeinschaft (DFG) (project MA4637/3-1), from the VITI Programme of the Helmholtz Association, by the MaX Center of Excellence funded by the EU through the H2020-EINFRA-2015-1 project: GA 676598 as well as support by the Joint Lab Virtual Materials Design (JLVMD). PR and SB acknowledge support by the Deutsche Forschungsgemeinschaft (DFG, German Research Foundation) under Germany's Excellence Strategy – Cluster of Excellence Matter and Light for Quantum Computing (ML4Q) EXC 2004/1 – 390534769.
This work was supported by computing time granted by the JARA Vergabegremium and provided on the JARA Partition part of the supercomputer CLAIX at RWTH Aachen University. P.R. would like to thank P. Mavropoulos, J. Br{\"o}der, and the AiiDA-team for fruitful discussions.


\section*{Author Contributions}

P. R. and S. B. conceived the project. P. R. and F. B. developed the AiiDA-KKR plugin. P. R. performed the calculations, developed the JuDiT webinterface for the data analysis, and wrote the initial manuscript. All authors contributed to the final manuscript.


\section*{Competing Interests}

The authors declare no competing financial or non-financial interests.


\appendix
\section*{Appendix}

\counterwithin{figure}{section}

\section{Impurity DOS for various transition metal impurities \lbl{app:1}}

\begin{figure*}[htb]
    \centering
    \includegraphics[width=17cm]{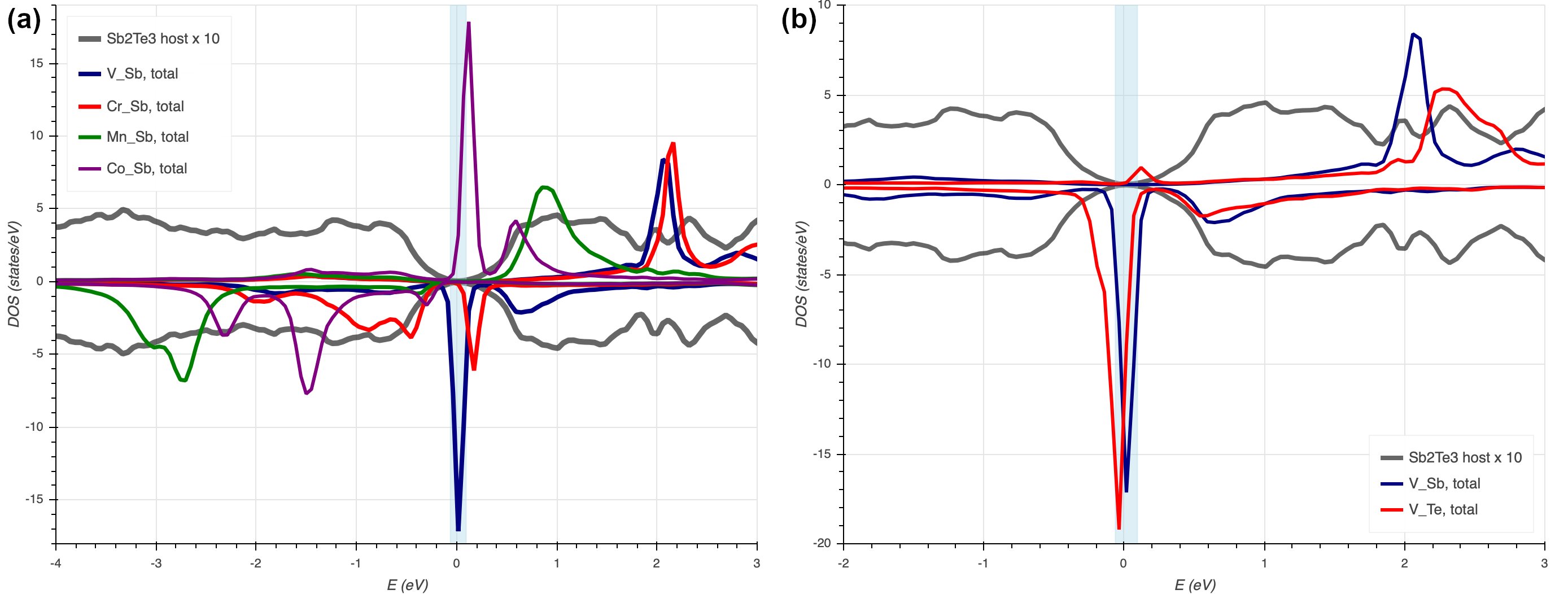}
    \caption{(a) DOS of magnetic $3d$ transition metal impurities embedded in the first Sb layer from the surface (see Fig.~\ref{fig:imp_results_overview}(c) for the position in the host crystal) and (b) DOS of V$_\mathrm{Sb}$ and V$_\mathrm{Te}$ in the outermost Sb and Te layers. A positive (negative) sign of the DOS indicates majority (minority) states of the impurity.}
    \label{fig:app1.1}
\end{figure*}

The DOS of $3d$ impurities in the outermost Sb layer (the same position as in Fig.~\ref{fig:imp_results_overview}(c)) 
is given in Fig.~\ref{fig:app1.1}(a). The increasing filling of the $d$ orbital discussed in the main text (Fig.~\ref{fig:layer_dependence}(a)) is seen. The smaller spin moment for V$_\mathrm{Te}$ than for V$_\mathrm{Sb}$ impurities can be traced back to the change in the impurity DOS induced by the differing surrounding host atoms.

\begin{figure}[htb]
    \centering
    \includegraphics[width=8.5cm]{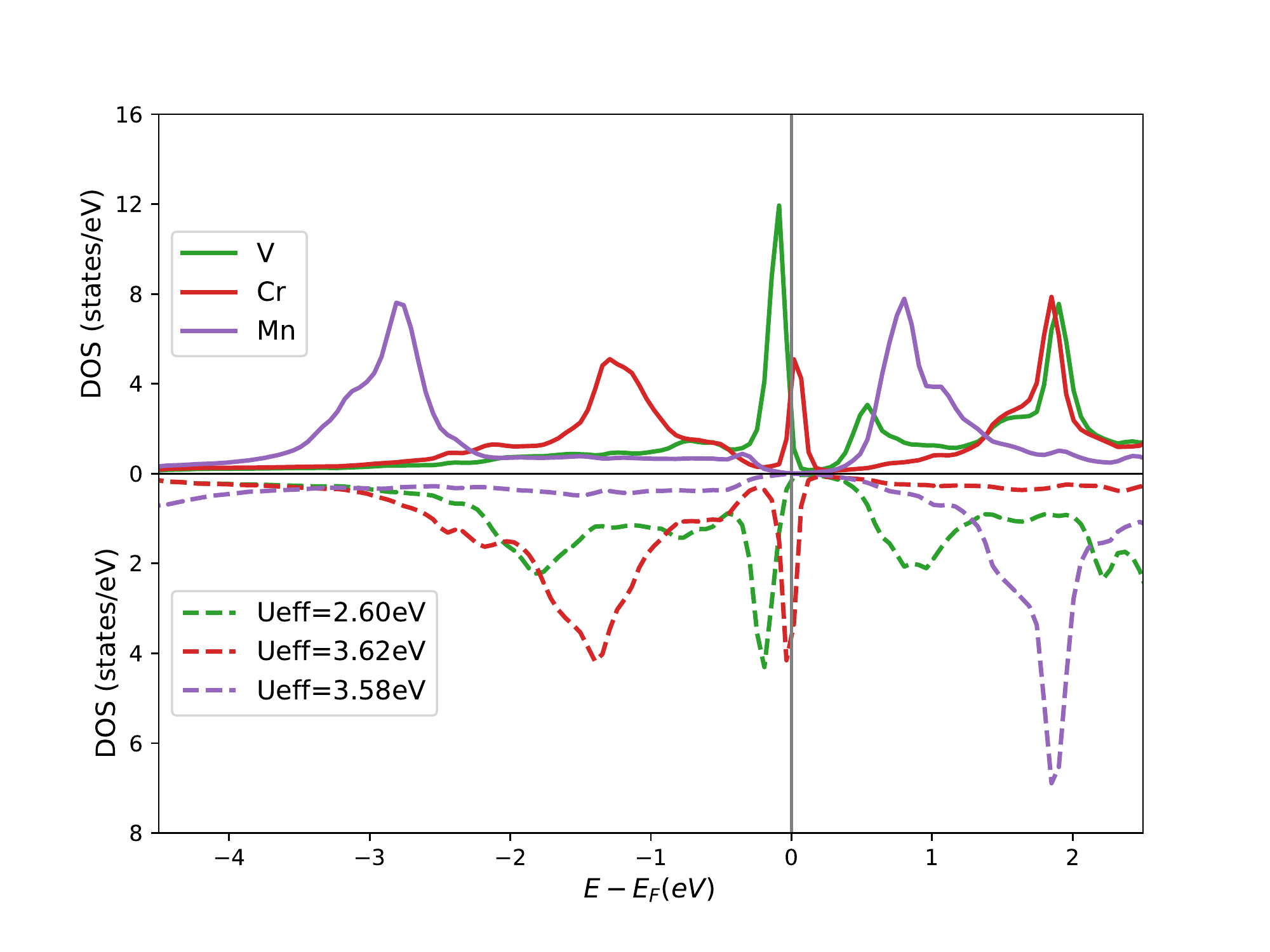}
    \caption{\addition{Comparison of the DOS without (top) and with (bottom) LDA+U corrections for $\mathrm{V}_{\mathrm{Sb}}$, $\mathrm{Cr}_{\mathrm{Sb}}$ and $\mathrm{Mn}_{\mathrm{Sb}}$ impurities located in the middle of the 6QL thick Sb$_2$Te$_3$ film. The values for the parametrization giving $U^{\mathrm{eff}}=U-J$ are taken from cRPA calculations of Ref.~\onlinecite{Sasioglu2011} and collected in Tab~\ref{tab:UJvals}.}}
    \label{fig:app1.2}
\end{figure}

\addition{
Additionally, the change in the DOS with including electron correlations by the LDA+U method is demonstrated in Fig.~\ref{fig:app1.2}. It can be seen that the on-site Coulombic repulsion leads to an enhanced exchange splitting that is shown exemplary in the impurity DOS for V, Cr and Mn defects. The U and J values of the LDA+U parametrization used in this work are given in Tab.~\ref{tab:UJvals}.
}

\begin{table*}[hbt]
    \centering
    \addition{
    \begin{tabular}{ccc|ccc|ccc}
$3d$ impurity & U (eV) & J (eV) & $4d$ impurity & U (eV) & J (eV) & $5d$ impurity & U (eV) & J (eV) \\\hline
Sc  &  2.3  & 0.40 & Y  & 1.6 & 0.29 & Lu & 1.4 & 0.30\\
Ti  &  3.0  & 0.50 & Zr & 2.4 & 0.35 & Hf & 2.0 & 0.37\\
V   &  3.2  & 0.60 & Nb & 2.6 & 0.45 & Ta & 2.3 & 0.46\\
Cr  &  4.3  & 0.68 & Mo & 2.6 & 0.51 & W  & 3.5 & 0.55\\
Mn  &  4.3  & 0.72 & Tc & 3.8 & 0.55 & Re & 3.6 & 0.57\\
Fe  &  3.7  & 0.72 & Ru & 4.2 & 0.59 & Os & 4.1 & 0.61\\
Co  &  4.3  & 0.77 & Rh & 4.1 & 0.60 & Ir & 3.8 & 0.62\\
Ni  &  3.8  & 0.78 & Pd & 3.7 & 0.60 & Pt & 3.6 & 0.62\\
Cu  &  5.5  & 0.85 & Ag & 4.8 & 0.64 & Au & 3.9 & 0.65\\
    \end{tabular}
    }
    \caption{\addition{U and J values used for the parametrization of the Coulomb repulsion in the LDA+U calculations. The values are extracted from Ref.~\onlinecite{Sasioglu2011}.}}
    \label{tab:UJvals}
\end{table*}

\section{\addition{Statistical analysis of impurity quantities in the JuDiT database}}

\begin{figure}[htb]
    \centering
    \includegraphics[width=8.2cm]{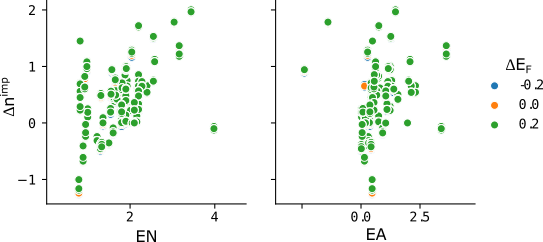}
    \caption{\addition{Pairwise relation between induced charge doping ($\Delta n^{\mathrm{imp}}$) and the impurity's Pauling electronegativity (left, EN) or electron affinity (right, EA). Each data point corresponds to a calculation in the JuDiT database. A Pearson correlation coefficient of $0.5$ (for EN) and $0.086$ (EA) is found for the two data relations indicating a causal relation only between $\Delta n^{\mathrm{imp}}$ and the impurity's EN but not with the EA.}}
    \label{fig:app2.1}
\end{figure}

\addition{
The large number of different impurities in the JuDiT database allows us to investigate the correlations between different physical properties of the defects. 
In Fig.~\ref{fig:app2.1} we show the correlation between the impurity's Pauli electronegativity (EN) and the induced charge doping. In contrast to the impurity's electron affinity we find that EN and $\Delta n^{\mathrm{imp}}$ are correlated. The corresponding values of the Pearson correlation coefficient is 0.5 which indicates the large spread of the data. This result highlights the need for \textit{ab initio} calculations which we collected in the JuDiT database.
}

\begin{figure*}[htb]
    \centering
    \includegraphics[width=17cm]{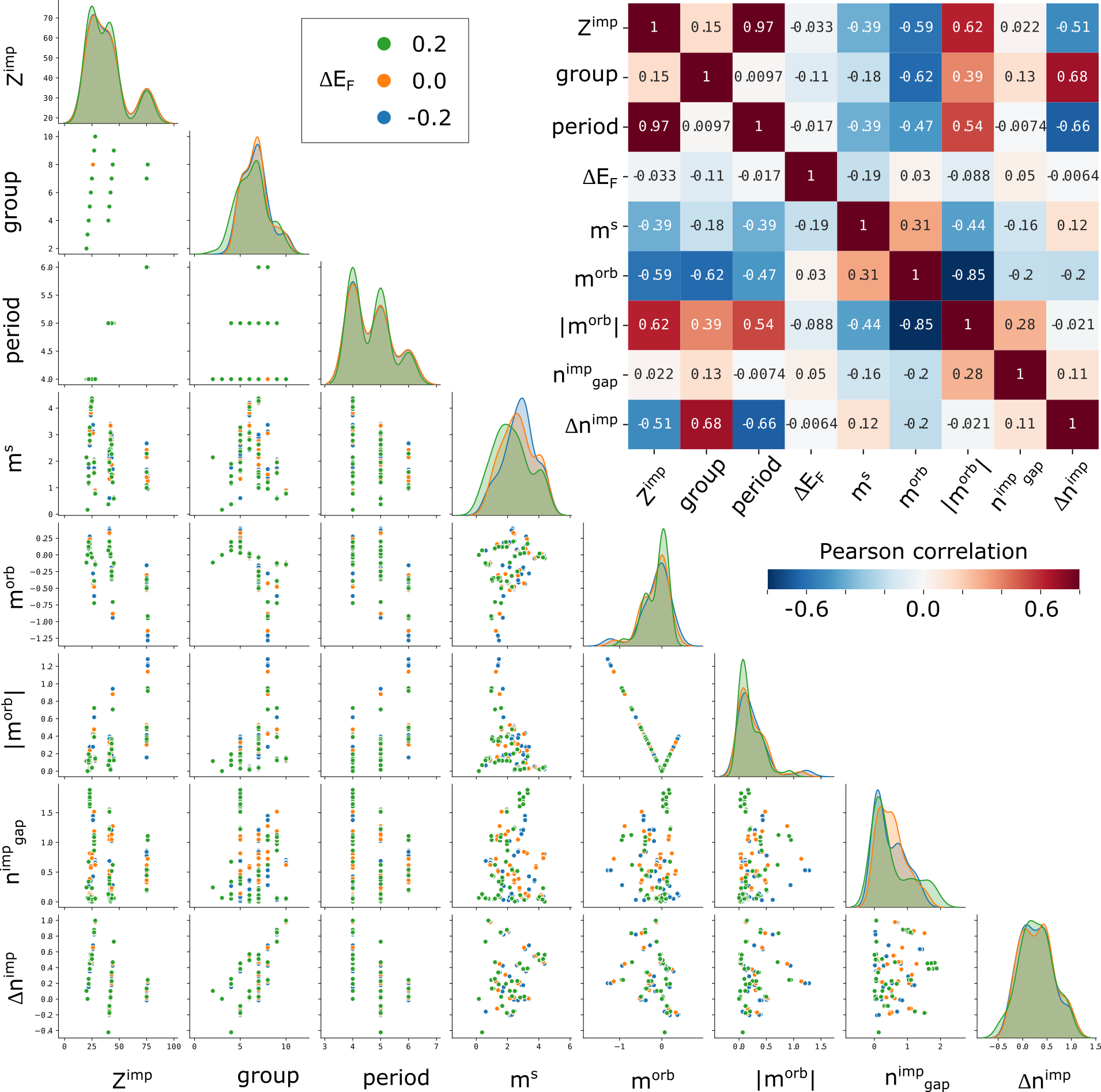}
    \caption{\addition{Pairwise plot of the data relations in the subgroup of magnetic defects without correlation effects in the JuDiT database. The plots on the diagonal shows the distribution of the quantities and the colors indicate the three values of the Fermi level.
    The corresponding Pearson correlation matrix is shown in the top right. Apart from the trivial correlation between $\mathrm{period}$ and $Z^{\mathrm{imp}}$, the strongest positive correlation is found for $(\mathrm{group},\Delta n^{\mathrm{imp}})$ and $(Z^{\mathrm{imp}},|m^{\mathrm{orb}}|)$. The strongest nontrivial negative correlation is seen for $(\mathrm{period},\Delta n^{\mathrm{imp}})$ and $(\mathrm{group},m^{\mathrm{orb}})$.}}
    \label{fig:app2.2}
\end{figure*}

\addition{
Furthermore we investigate the pairwise correlations between different impurity properties for the defects in the JuDiT database.
This is shown in Fig.~\ref{fig:app2.2} where we see that the charge doping $\Delta n^{\mathrm{imp}}$ correlates with the impurity's group index of the periodic table and that the size of the orbital moment scales with the nuclear charge.
}

\begin{figure}[htb]
    \centering
    \includegraphics[width=8.5cm]{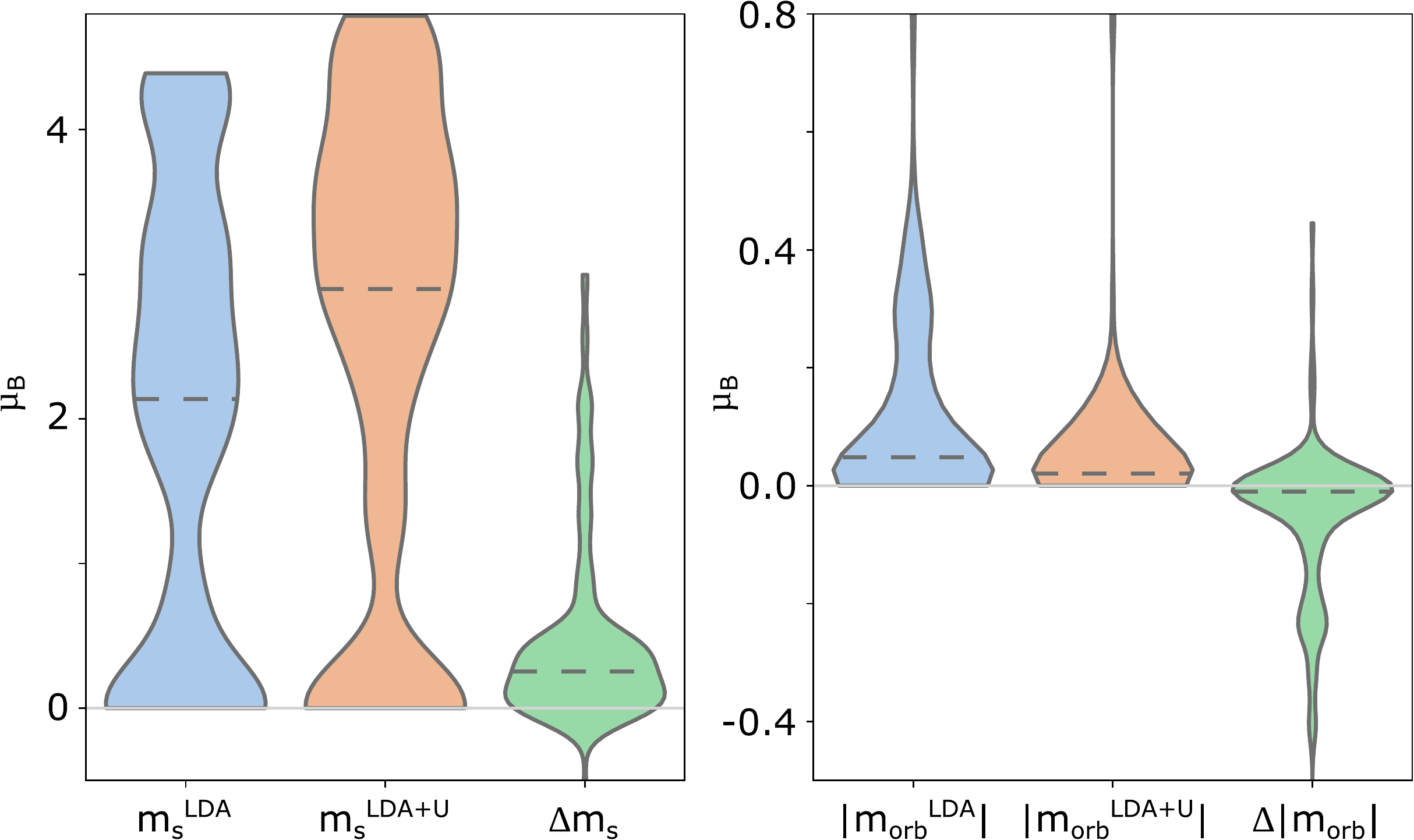}
    \caption{\addition{Spin (left) and orbital (right) moments of the 153 $d$ impurities in the JuDiT database for which LDA+U calculations are included. The violin plots indicate the distribution of the size of the magnetic moments in LDA (blue) and LDA+U (orange) as well as the differences $\Delta m_s=m_s^{\mathrm{LDA}} -m_s^{\mathrm{LDA+U}}$ (green). The dashed lines indicate the median values of the distributions.}}
    \label{fig:app2.3}
\end{figure}

\addition{
The change in the magnetic properties in transition metal impurities with the inclusion of correlation effects in the LDA+U method is summarized in Fig.~\ref{fig:app2.3}. 
We see that the median values of the spin (orbital) moment increases (decreases) with including on-site repulsion. This goes hand in hand with the enhanced exchange splitting in the impurity DOS that was discussed in Fig.~\ref{fig:app1.2}. A larger number of non-magnetic defects remain non-magnetic which is seen in the peak at $m_{s}=0$ of the distribution plots.
}

\end{document}